\documentclass[useAMS,usenatbib,twocolumn]{mnras}
\usepackage{times}
\usepackage{graphicx}
\usepackage{lscape}
\usepackage{ amssymb }
\def\etal{{\it et al.}}

\title[A deceleration search for pulsations in SGRB plateaus]{A deceleration search for magnetar pulsations in the X-ray plateaus of Short GRBs}
\author[A. Rowlinson \etal]{A. Rowlinson$^{1,2}$\thanks{E-mail: b.a.rowlinson@uva.nl}, A. Patruno$^{3,2}$, P.T. O'Brien$^{4}$\\ 
$^{1}$Anton Pannekoek Institute, University of Amsterdam, Postbus 94249, 1090 GE Amsterdam, The Netherlands\\
$^{2}$ASTRON, the Netherlands Institute for Radio Astronomy, Postbus 2, 7990 AA Dwingeloo, The Netherlands\\
$^{3}$Leiden Observatory, Leiden University, P.O. Box 9513, NL-2300 RA Leiden, The Netherlands\\
$^{4}$Department of Physics \& Astronomy,University of Leicester, University Road, Leicester, LE1 7RH, UK
}

\begin{document}

\pagerange{\pageref{firstpage}--\pageref{lastpage}} \pubyear{000}
\maketitle            

\label{firstpage}

\begin{abstract}

A newly formed magnetar has been proposed as the central engine of short GRBs to explain on-going energy injection giving observed plateau phases in the X-ray lightcurves. These rapidly spinning magnetars may be capable of emitting pulsed emission comparable to known pulsars and magnetars. In this paper we show that, if present, a periodic signal would be detectable during the plateau phases observed using the {\it Swift} X-Ray Telescope recording data in Window Timing mode. We conduct a targeted deceleration search for a periodic signal from a newly formed magnetar in 2 {\it Swift} short GRBs and rule out any periodic signals in the frequency band 10--285 Hz to $\approx$15--30\% rms. These results demonstrate that we would able to detect pulsations from the magnetar central engine of short GRBs if they contribute to 15-30\% of the total emission. We consider these constraints in the context of the potential emission mechanisms. The non-detection is consistent with the emission being reprocessed in the surrounding environment or with the rotation axis being highly aligned with the observing angle. As the emission may be reprocessed, the expected periodic emission may only constitute a few percent of the total emission and be undetectable in our observations. Applying this strategy to future observations of the plateau phases with more sensitive X-ray telescopes may lead to the detection of the periodic signal.

\end{abstract}


\section{Introduction}

\begin{table*}
\begin{tabular}{|c|c|c|c|c|c|c|c|c|c|c|}
\hline
GRB             & Redshift & $T_{90}$      & B$_{15}$      & P$_{-3}$   & Restframe Collapse Time & WT observation time \\
                &          & (s)           & ($10^{15}$ G) & (ms)       & (s)                     & (s) \\
\hline
090510$^{(1)}$  & 0.9      & 0.3$\pm$0.1   & 5.06$^{+0.27}_{-0.23}$ & 1.86$^{+0.04}_{-0.03}$ & -   & 94--294 \\
090515$^{(2)}$  & 0.403   & 0.04$\pm$0.02 & 14.15$^{+2.39}_{-2.46}$ & 4.16$^{+0.18}_{-0.22}$ & 214 & 70--195 \\
090515$^{(2)}$  & 0.657   & 0.04$\pm$0.02 & 9.37$^{+1.30}_{-1.30}$ & 2.53$^{+0.1}_{-0.1}$ & 181 & 70--195 \\
\hline
\end{tabular}
\caption{Basic properties of the sample GRBs, for the hostless GRB 090515 we utilise the redshift values from the two most likely host galaxies \citep{berger2010}. Magnetar parameters are obtained from the magnetar fits in \citet{rowlinson2013} and \citet{postigo2013}, assuming that the magnetar emission is isotropic and 100\% efficient (see text for more details). If the magnetar is unstable, the restframe collapse time is provided. Additionally, we give the observed frame WT observation start and end times relative to the trigger time.}
$^{(1)}$ \cite{ukwatta2009, depasquale2010, mcbreen2010}
$^{(2)}$ \cite{barthelmy2009, rowlinson2010, berger2010}
\label{sample}
\end{table*}

The launch of the {\it Swift} satellite \citep{gehrels2004} has led to a reformation in our understanding of early afterglow emission from gamma-ray bursts (GRBs). Particularly, {\it Swift} highlighted that central engine activity is often long lived, powering flares and plateaus \citep{nousek2006,zhang2006,obrien2006}. Prolonged central engine activity is often explained as ongoing accretion onto the newly formed black hole (BH) following the collapse of a massive star \citep[e.g.][]{macfadyen2001}.

For Short GRBs (SGRBs), typically with durations of $T_{90} \le 2$ s\footnote{Though we cannot rely upon prompt emission alone to unambiguously identify SGRBs \citep[e.g.][]{bromberg2013, qin2013}}\citep{kouveliotou1993}, prolonged accretion is not expected within the standard progenitor model. They are thought to originate from the merger of a compact binary system constituting of two neutron stars (NSs) or a NS and a BH \citep{lattimer1976,eichler1989,narayan1992}. In this model the accretion is expected to end within $\sim 2$ s \citep[e.g.][]{rezzolla2011} powering the prompt gamma-ray emission. Possible late time accretion of material on highly eccentric orbits could lead to flares in the X-ray lightcurve but cannot power prolonged plateau phase \citep[e.g.][]{rosswog2013}. However, studies of SGRB X-ray lightcurves has shown that there is evidence of plateau phases signifying prolonged energy injection that cannot be explained by this theory \citep{rowlinson2010, rowlinson2013, lu2014}. 

An alternative model is that the central engine of GRBs is a newly formed millisecond pulsar with a high magnetic field and sufficient rotational energy to prevent gravitational collapse \citep[referred to as a magnetar, ][]{usov1992,duncan1992,dai1998a,dai1998b,zhang2001}. The magnetar can be formed in a variety of ways; during the collapse of a massive star \citep[e.g.][]{metzger2011}, via accretion induced collapse of a NS or a white dwarf \citep[e.g.][]{usov1992} or the merger of two NSs \citep{dai1998a,dai2006,yu2007}. This model predicts a plateau phase in the X-ray lightcurves originating from dipole emission from the rapidly spinning down magnetar \citep[assuming constant radiative efficiency, ][]{zhang2001}. As the magnetar spins down, the plateau slowly turns over to a powerlaw decline. If the newly formed magentar is unstable (i.e. the mass supported by its rapid rotation is greater than the maximum allowed mass of a NS), then it will reach a critical point at which it is unable to support itself and will instead collapse to form a BH. At that point, the energy injection is rapidly turned off leading to a steep decay phase in the X-ray lightcurve rather than a shallow decay phase \citep{troja2007,lyons2010,rowlinson2010,rowlinson2013}.

This model has been fitted to large samples of Long GRBs \citep[LGRBs; e.g.][]{lyons2010,dallosso2011,bernardini2012,lu2014,yi2014}, all {\it Swift} SGRBs with sufficient X-ray observations \citep{rowlinson2010,rowlinson2013, postigo2013, lu2014} and has been proposed to explain energy injection in the class of SGRBs with extended emission \citep[e.g.][]{metzger2008,bucciantini2012, gompertz2013, gompertz2014,gibson2017}. The fitted magnetar parameters for all of these candidates are consistent with the expected values for newly formed magnetars, although there is no conclusive proof to date that magnetars are the central engines. \cite{rowlinson2013} suggested that the next generation gravitational wave detectors may be able to provide this proof, as a newly formed magnetar has been predicted to produce an additional gravitational wave signal following the initial inspiral signal \citep{corsi2009,giacomazzo2011a,melatos2014,dallosso2015,lasky2016}, however the expected detection rates are very low.

Alternatively, if a magnetar is the central engine powering GRBs, we might expect to see periodic features in the emission. Known magnetars have clear periodic signals in their emission caused by their rotation periods \citep[e.g.][]{mazets1979,kouveliotou1998}. The X-ray pulsations typically contribute to 30\% of the signal, with a range of 10--80\% \citep{israel1999,kargaltsev2012,kaspi2017}. There is an energy dependence on the pulsed fraction of the signal, where low energies tend to have smaller pulsed frations \citep{vogel2014}. Detection of a periodic signal during the plateau phase in the X-ray lightcurve would provide excellent supporting evidence for the magnetar central engine model. There have been searches for a periodic signal in the prompt emission of GRBs with a number of instruments with no success, for example: BATSE GRBs \citep[Burst And Transient Source Experiment;][]{deng1997}, INTEGRAL GRBs \citep[INTErnational Gamma-Ray Astrophysics Laboratory;][]{ryde2003}, GRB 051103 \citep[an extragalactic Soft Gamma-ray Repeater giant flare candidate detected by the Inter Planetary Network;][]{hurley2010} and BAT GRBs \citep[Burst Alert Telescope;][]{cenko2010,deluca2010,guidorzi2012}. \cite{dichiara2013} searched the prompt emission of a number of short GRBs for evidence of a precessing jet \citep[predicted by ][]{stone2013}. However, these searches typically target the prompt emission and have not probed the regime where we might expect periodic signals from a magnetar central engine (i.e. during the plateau phase). Only two GRBs have been searched for periodic emission during the X-ray observations when the magnetar central engine may dominate the emission, GRB 060218 \citep{mirabal2010} and GRB 090709A \citep{mirabal2009, deluca2010}. The prompt emission of GRB 090709A possibly showed evidence of a periodic signal \citep{markwardt2009, golenetskii2009, gotz2009, ohno2009}, however this was ruled out with more careful analysis of the prompt data from BAT, XRT (X-ray Telescope) and {\sc XMM} (X-ray Multi-mirror Mission) observations of the X-ray afterglow \citep{cenko2010,deluca2010}. However, in the majority of these studies, the authors have targeted a constant spin period whereas a magnetar central engine is expected to have a rapidly decelerating spin period which would be very difficult to detect in standard searches for periodic signals. \cite{dichiara2013} did conduct a deceleration search, however they were targeting signals in the prompt emission where we do not expect the signal from a spinning down magnetar.
 
In this paper, we present the first targeted deceleration search for a periodic signal associated with a spinning down magnetar central engine. For a successful periodicity search we require:
\begin{itemize}
    \item A GRB which is not in a high density environment or have a progenitor which may have blown off a large amount of material, as this could lead to reprocessing of the emission which may dilute the periodic signal.
    \item A plateau phase showing evidence of energy injection within the X-ray observations. The magnetar component should dominate the lightcurve in order to get the largest periodic signal, so we need GRBs which have a minimal standard afterglow component.
	\item Window Timing \citep[WT;][]{burrows2005} mode observations covering part of the plateau phase. WT mode provides the timing resolution required for a millisecond periodicity search.
	\item A good redshift constraint.
\end{itemize}
We propose that SGRBs provide the ideal dataset for this analysis as they are expected to occur in low density environments and typically have a faint afterglow. From the analysis in \cite{rowlinson2013}, we identified two, unambiguously short, SGRBs which satisfied these criteria: GRB 090510 and GRB 090515 \citep[most likely to originate from one of 2 galaxies at redshifts of 0.403 and 0.657;][]{berger2010}. Section 2 describes the periodic signals predicted from the magnetar central engines that are consistent with the X-ray lightcurves of these SGRBs. In Section 3 we describe the periodicity search conducted and provide the results, while Section 4 discusses the theoretical implications of our observations and the likelihood of the production of a detectable signal.

\section{Periodic signal predictions}

The magnetar spin period and spin-down rate are analytically predictable using the dipole radiation model which is fitted to the X-ray lightcurves of the SGRBs. The initial magnetic field strength and spin period are given by \citep{zhang2001}: 

\begin{eqnarray}
B^{2}_{p,15}=4.2025 I_{45}^{2}R^{-6}_{6}L_{0,49}^{-1}T_{em,3}^{-2}\left(\frac{\epsilon}{1-\cos\theta}\right),   \label{b^2}   \\
P^{2}_{0,-3}=2.05 I_{45}L_{0,49}^{-1}T_{em,3}^{-1}\left(\frac{\epsilon}{1-\cos\theta}\right).     \label{p^2}
\end{eqnarray}

Where $B_{15}$ is the magnetic field strength of the newly born magnetar in $10^{15}$ G, $P_{0,-3}$ is the initial spin period of the magnetar in ms, $I_{45}\sim M_{1.4}R_{6}^{2}$ is the moment of inertia of a NS where $I = 10^{45} {\rm~g~cm^{2}~} I_{45}$, $R_{6}$ is the radius of the magnetar in $10^6$ cm, $M_{1.4}$ is the mass of the magnetar in $1.4 M_{\odot}$, $L_{0,49}$ is the plateau luminosity in $10^{45}$ erg s$^{-1}$ and $T_{em,3}$ is the plateau duration in $10^3$ s. In both equations, we also include the dependence on the beaming angle ($\theta$) and the efficiency in conversion of the rotational energy into the observed X-ray emission ($\epsilon$). Additionally, we do not know the mass or radius of the newly formed magnetar. The mass of the newly formed magnetar is expected to be $1 \le M_{1.4} \le 1.5$ therefore, as $P_{0,-3} \propto M_{1.4}^{0.5}$, the spin period is only expected to vary by $\sim20$\% which is not significant in comparison to the other uncertainties caused by efficiencies and beaming. Magnetars may be formed with radii up to $\sim 30$ km \citep{ott2006}, however it is expected that they will stabilise at a typical NS radius of $R_{6} \sim 1$ within the first few seconds \citep{metzger2011}. Therefore, in this paper we assume $M_{1.4}=1.5$, as the newly formed magnetar is most likely to be a massive neutron star, and $R_{6}=1$.

The magnetar emission is assumed to be isotropic and 100\% efficient for fitting purposes, however it is important to note that this is an idealised situation and changes to this assumption can cause significant differences in the output values for $B_{15}$ and $P_{0,-3}$ \citep[see discussion in][]{rowlinson2010, rowlinson2013}. We account for these uncertainties later in this section. Using the method described in \cite{rowlinson2013}, the observed 0.3--10 keV lightcurves of the SGRBs were converted into restframe 1--10000 keV lightcurves and fitted using the magnetar central engine plateau model, given by equations \ref{p^2} and \ref{b^2}, and the parameters for each of these GRBs are provided in Table \ref{sample}. These fits determine the initial spin period of the magnetar, however the magnetar is rapidly spinning down so we also need to predict the spin down evolution. To do this we can use, from \cite{piro2011},  
\begin{eqnarray}
\frac{d\Omega}{dt} = \frac{N_{\rm dip}}{I}  \label{piro1}\\
N_{\rm dip}=-1.5\times10^{45}\mu_{33}^{2}P^{-3}_{-3} \label{piro2}
\end{eqnarray}
assuming there is no ongoing accretion. Where $\mu$ is the dipole magnetic moment, $\mu_{33}=B_{15}R_{6}^{3}=10^{33}{\rm ~G~cm^{3}}~\mu$, $\Omega=\frac{2\pi}{P_{-3}}$ is the angular velocity and $N_{\rm dip}$ is the torque from the dipole emission \citep{piro2011}. We note \cite{bucciantini2006} derive a more complex torque from dipole emission, taking into account open magnetic flux tubes in an accreting magnetar system, however the accretion is expected to have ended prior to the emission we observe and this additional complexity not required. By substitution of Equations \ref{b^2} and \ref{p^2} into Equations \ref{piro1} and \ref{piro2} followed by integration, we can predict that the spin period evolution with time can be described by: 
\begin{eqnarray}
\nu=\left(5\times10^{-7}xt + 10^{-6}P_{0,-3}^{2}\right)^{-\frac{1}{2}}~{\rm s}, \label{nu_t}
\end{eqnarray}
where $\nu \equiv 1/P$ and
\begin{eqnarray}
x=\frac{B_{15}^{2}R_{6}^{4}}{2\pi M_{1.4}}.  \label{const}
\end{eqnarray}
By differentiation we can determine the spin down rate to be given by:
\begin{eqnarray}
\dot\nu=-5\times10^{-7}\frac{x}{2}\nu^{3}~{\rm Hz\,s}^{-1} \label{nudot_t}
\end{eqnarray}
Additionally, we assume that the magnetar is spinning down purely via dipole radiation so the relationship between the spin and its spin down properties are well defined using the breaking index:
\begin{eqnarray}
n = \frac{\nu \ddot\nu}{\dot\nu^{2}} ~(= 3 {\rm ~ for~dipole~spindown}).   \label{breaking}
\end{eqnarray}
This assumption is intrinsic to the magnetar model typically fitted to the X-ray plateaus \citep{zhang2001}, however known young pulsars are known to be spinning down differently to this, with braking indices $n<3$, \citep[e.g.][]{manchester1985, lyne1993, camilo2000, livingstone2007, espinoza2011}. Recently, \cite{lasky2017} extended the magnetar model to use the late time decay slope to constrain the spin down of a magnetar central engine in two SGRBs. One of their sample is fitted with $n=2.6 \pm 0.1$ (GRB 140903A), consistent with the observed $n<3$ braking indices in millisecond magnetars. The other, GRB 130603B, has $n=3.0 \pm 0.1$ as expected for dipole radiation. Therefore, the assumption of pure dipole radiation is likely to be consistent for at least some of the magnetar engines fitted in the SGRB sample but likely not all. Unfortunately, neither GRB in \cite{lasky2017} have sufficient WT mode data to be included in our sample. However, it is promising that in the future we may be able to directly measure the braking index for SGRBs and, combined with the required WT mode data, obtain a much deeper constraint on periodic emission. In this paper, we consider the impact of different braking indices and this issue will be discussed further in Section 3. 

\begin{figure}
\centering
\includegraphics[width=.45\textwidth]{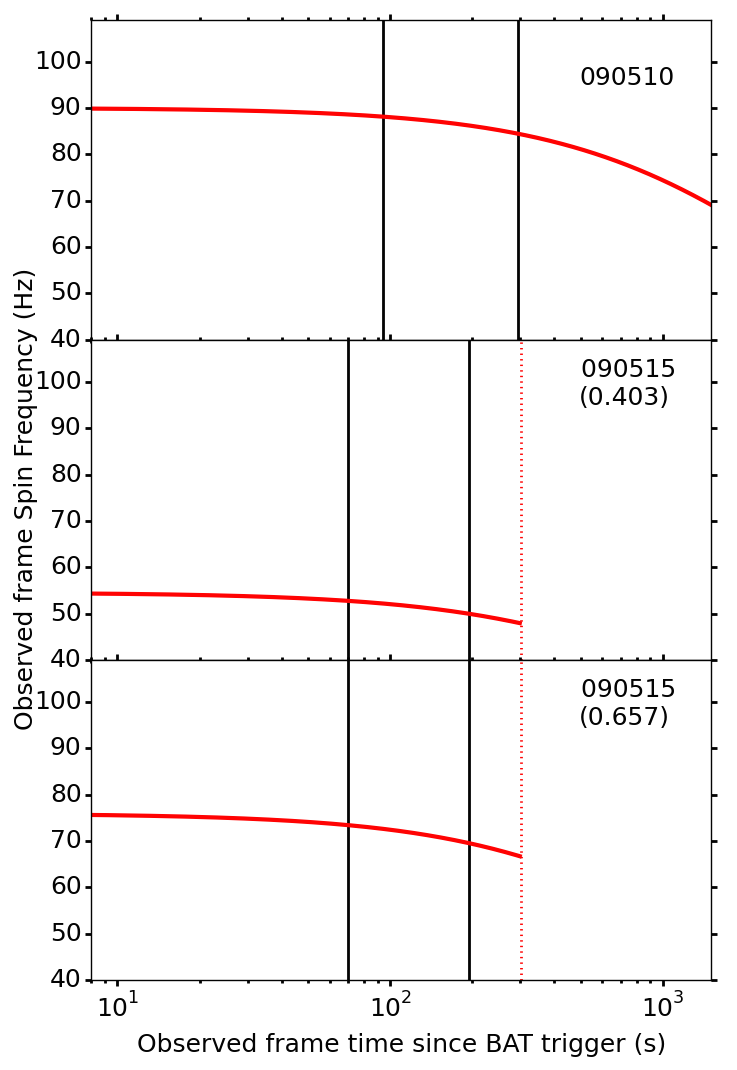}
\caption{The spin frequency of the magnetar evolves with time since the formation of the magnetar (assuming it forms at the time of the BAT trigger), here we plot the observer frame spin frequency for each of the GRBs in the sample. The black solid lines show the start and end of the WT mode observations and the red dotted line marks the time at which the magnetar collapses to form a BH. This plot assumes a beaming angle of $10^{\circ}$ and efficiency of 10\%.}
\label{theory}
\end{figure}

Using Equations \ref{nu_t}, \ref{nudot_t} and \ref{breaking} alongside the magnetic field strengths and restframe spin periods obtained, we can describe how the spin frequency of the newly formed magnetar evolves with time. The restframe spin frequencies are then converted into observed frame spin frequencies that we might expect to detect from the timing analysis conducted in Section 3. In Figure \ref{theory} we show how the spin frequency evolves during the WT mode observation for each of the SGRBs. However, from these plots it is clear that the spin frequency can decrease significantly from the start of the WT observation to the end of the WT observation so any periodicity searches will need to account for this rapid spin down. 

\begin{figure}
\centering
\includegraphics[width=0.45\textwidth]{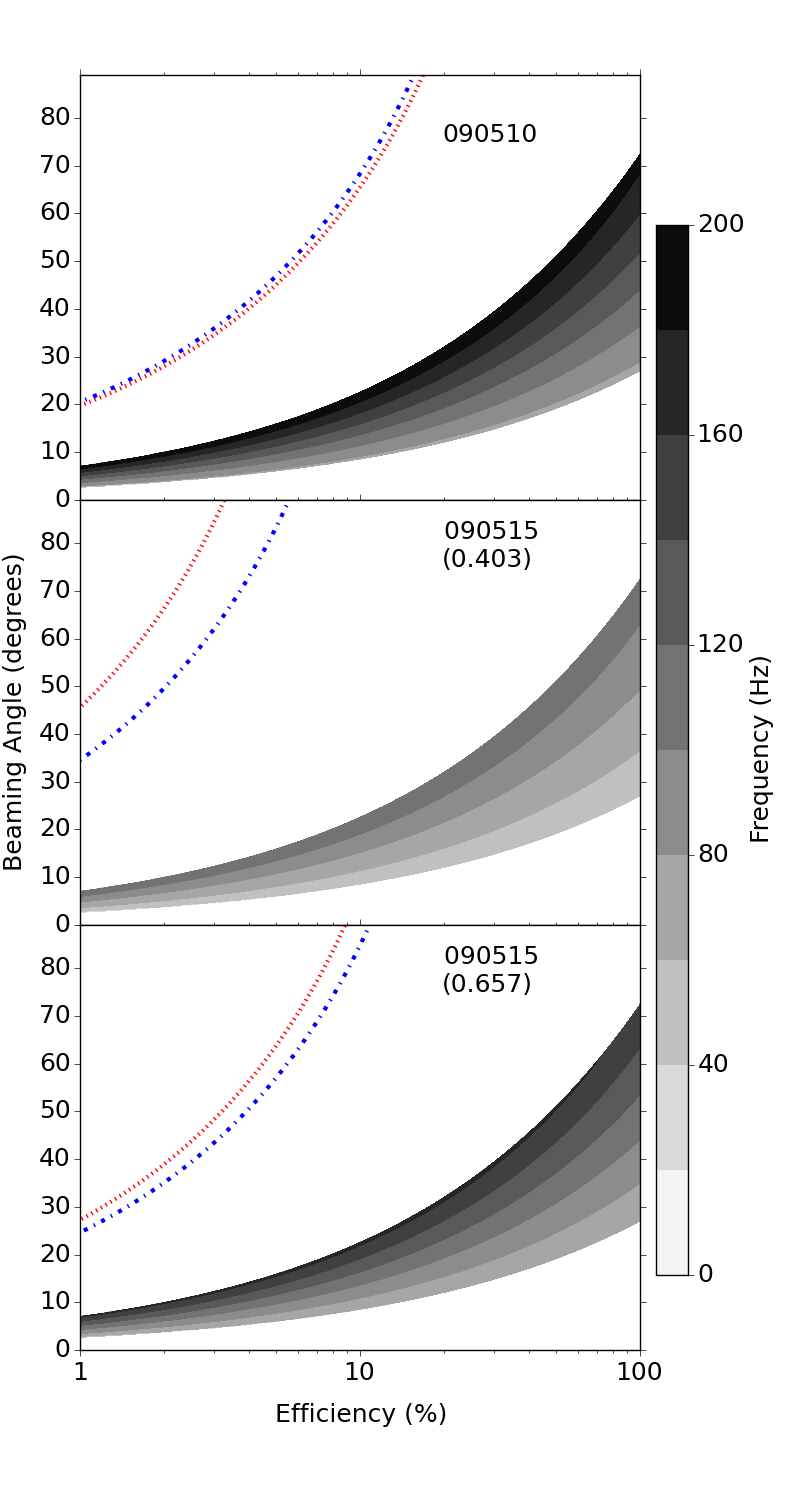}
\caption{The range of observer frame spin frequencies (at the start of the WT mode observation) for each GRB is plotted as a function of the beaming angle of the emission and the efficiency of converting the rotational energy to the observed X-ray emission. The blue dot-dash line represents the WT mode resolution and all combinations in the white portion of each plot would be undetectable. The red dotted line shows the spin break-up frequency of a 1.4$M_{\odot}$ NS, above which no stable NSs can be formed. The black dashed lines represent the 50\% probability contours from \citet{rowlinson2014}, which constrain the efficiency and beaming angles to values compatible with the observed GRB sample.}
\label{eff_beam}
\end{figure}

As previously stated, the efficiency in converting the rotational energy into the observed plateau and the beaming angle of the emission have a significant impact on the spin periods predicted. However, both of these are currently unknown; here efficiencies are assumed to lie within the range 1--100\% while jet opening angles for SGRBs are thought to range from 1--20 degrees or more \citep[e.g.][]{popham1999, ruffert1999, aloy2005, rosswog2005, rezzolla2011}. Figure \ref{eff_beam} shows the periodicity at the start of the WT mode observation for each of the GRBs as a function of both the efficiency and beaming of the observation. The region above the blue dash-dot line illustrates the region that cannot be probed using the observations due to the timing resolution of the {\it Swift} WT mode observations. We also show the spin break up frequency of a 1.4 $M_{\odot}$ NS in the observer frame for each of the GRBs (red dotted line) above which no NSs can exist. \cite{rowlinson2014} showed that the observed correlation between the plateau luminosity and duration for GRBs \citep[e.g.][]{dainotti2008} can be used to tightly constrain the efficiency and beaming angle of the emission from the magnetar central engine. A probability contour plot was produced by this analysis, providing the probability that the magnetar model is consistent with the observed dataset as a function of different beaming angles and efficiencies. We use the 50\% probability contours from the analysis of \cite{rowlinson2014} to reject regions of the beaming and efficiency parameter space, thus more tightly constraining the properties of the magnetar. The upper and lower 50\% contours are well fitted with simple exponential equations and we use these fits to incorporate the allowed region of the parameter space into the modelling of the periodic signal. All combinations of beaming angles and efficiencies that do not lie within these contours are excluded from the modelling. After applying these constraints, we note that all of the expected spin periods for the SGRBs lie within the detectable range for the WT mode data. Using the allowed spin periods, we extract the range of values for $\nu$, $\dot\nu$ and $\ddot\nu$ that we want to probe for each GRB, provided in Table \ref{limits}. Although these numbers are strongly related (see equations \ref{nu_t}--\ref{breaking}), we search the entire region of this parameter space for simplicity.

\begin{table*}
\begin{tabular}{|c|c|c|c|c|c|c|c|c|c|c|}
\hline
GRB             & $\nu$      & $\dot\nu$     & $\ddot\nu$   \\
                & (Hz)            & (-1 $\times10^{-3}$ Hz s$^{-1}$) & ($\times10^{-6}$ Hz s$^{-2}$)  \\ 
\hline
090510          & 74--198    & 8.7--23 & 3.0--8.2 \\
090515 (0.403)  & 44--118    & 14--39 & 14--38 \\
090515 (0.657)  & 62--165    & 17--46 & 14--38 \\
\hline
\end{tabular}
\caption{The parameter space of $\nu$, $\dot\nu$ and $\ddot\nu$ in which we want to search for a periodic signal for each GRB. We provide two search regimes for 090515, corresponding to the two possible redshifts for this burst.}
\label{limits}
\end{table*}

This analysis has shown that a periodic signal resulting from a magnetar spinning down via dipole radiation would be detectable by {\it Swift} XRT in WT mode observations for these SGRBs and reasonable combinations of the efficiency and beaming angle. Note this analysis assumes that the plateau emission contains a highly pulsed component which is detectable above the continuum emission, in Section 4 we discuss the likelihood of this.

\section{Timing Analysis Method and Results}

\subsection{X-Ray Data}

In view of our preceding discussion, we analyzed two targeted \textit{Swift}/XRT observations on the two GRBs previously discussed. We used only data recorded with a time resolution of 1.766 ms (WT-mode) and we extracted the source counts from a circular region of radius 30 arcsec centered on the brightest pixel and in the energy range 0.5--10 keV. The background was calculated from a similar extraction region placed as far as possible from the source location. 
A summary of the total length of the observations, the total number of photons, the background count-rate and the observation ID are summarized in Table~\ref{tab:obs}.

\begin{table}
\begin{tabular}{|c|c|c|c|c|}
\hline
GRB             & ObsID & Exposure  & Nr. Photons & Bkg count rate\\
  	 			& 		& (s)		& 			& (ct/s)\\
\hline
090510   & 351588000 & 172 & 800 & 0.06\\
090515   & 352108000 & 134 & 916 & 0.16\\
\hline
\end{tabular}
\caption{Summary of \textit{Swift}/XRT observations used in the pulse search.}
\label{tab:obs}
\end{table}

\subsection{Simple Periodicity Search}

The first type of pulse search we adopted is the simplest one and is based on a by-eye inspection of power spectra of different length. We calculated Fourier transforms with length between 4 s and 128 s with no background subtraction and/or dead time correction applied prior to the calculation. Then we averaged each power spectrum by Leahy-normalizing them by subtracting a Poissonian (counting) noise level incorporating dead-time effects as explained in \citet{zha95}.

We first looked for candidate pulsations with a power exceeding a threshold power of 30, which would correspond to a 3-$\sigma$ detection (single trial). We then produced dynamical power spectra of 4-s length each and looked for patterns in the peak powers. In neither case we had a candidate to follow up. 
If we assume that the pulse power remains in one Fourier frequency bin during the entire observation, then we can place upper limits on the root-mean-square (rms) pulse amplitude under the assumption that the power spectrum contains only white (counting) noise~\citep{van88}: 
\begin{equation}
{\rm\,rms} = [2(S/N)\cdot(S+B)]^{1/2}\,S^{-1}T_{\rm obs}^{-1/2}
\end{equation}
where $S$ and $B$ are the signal and background count rates, respectively, $S/N$ is the target signal-to-noise of the pulsations (i.e., the single trial significance) and $T_{\rm obs}$ is the length of the power spectrum (in seconds). 
Upper limits for a $S/N\approx\,5$ are of the order of 10\% rms for both observations. When looking at the 4-s long dynamical power spectra, we would have detected a signal with a $S/N\approx3$ if the rms amplitude of the pulsations had been in excess of approximately 50\%. 

We caution that since we expect a very rapid drift of the pulse frequency over time the power will spread across multiple bins. Therefore our assumption of having the power in one Fourier frequency bin breaks down and the aforementioned 10\% upper limits become unrealistic. The amount of bins over which the power spreads depends on the deceleration of the pulsar. 

\subsection{Deceleration Search}

As a first approximation we can consider a neutron star decelerating at a constant rate. The maximum number of bins $z_{\rm max}$ over which the spin frequency power will spread is thus~\citep{ran02}:
\begin{equation}
z_{\rm max} = \frac{a_{\rm max} T_{\rm obs}^2 N_{\rm harm}\,\nu}{c}
\end{equation}
where $a_{\rm max}$ is the maximum allowed (radial) deceleration, $N_{\rm harm}$ is the harmonic number and $c$ is the speed of light. 
The acceleration can be calculated from our estimated $\dot{\nu}$ in the preceding sections. Since the maximum $\dot{\nu}$ is of the order of $-0.01\rm\,Hz\,s^{-1}$, our maximum acceleration would 
give a drift of the order of $50,000\rm\,m\,s^{-2}$ and a maximum number of bins of the order of a few hundreds. 

To begin with, we performed a deceleration search with the software PRESTO (v.17Mar15) on our \textit{Swift}/XRT time series~\citep{ran02}. The search uses matched filtering techniques to add power of a drifting spin frequency under the assumption that the drift is approximately constant in time (i.e., there is a constant deceleration). 
The search was carried for frequencies in the range 10 to 283 Hz (i.e., our Nyquist frequency) and for $z_{\rm max}=800$. We searched pulsations under the assumptions that no harmonic content was present in the data, which is a good assumption if the expected pulse emission patters is nearly sinusoidal (as is the case for a Lambertian emitter like a hot spot). The significance is calculated by looking at the power returned by the matched filtering technique and then it is transformed into a false alarm probability from a chi-square distribution. No candidate above 3$\sigma$ was found in any of the GRB used. 

\subsection{Sensitivity and Upper Limits}

To determine the sensitivity of our search we performed a set of Monte-Carlo simulations where we generated simulated time series having the same sampling time, number of photons (following a Poisson distribution) and duration of the original \textit{Swift}/XRT time series. The simulated time series contain an injected sinusoidal signal whose phase evolves in time. The time evolution is described in terms of a frequency, frequency derivative and braking-index, whose values cover a 3D grid (see Table~\ref{tab:grid}). The deceleration search is then applied to the time series. The procedure is then repeated by increasing the amplitude of the signal from a minimum of $10\%$ rms in steps of 2\% up to 50\% rms.

\begin{table}
\begin{tabular}{|c|c|c|c|}
\hline
Parameter & Initial Value & Step size & Number of Steps\\
\hline
$\nu$ (Hz)  & 10 & 10 & 27\\
$\dot{\nu}$ ($\rm\,Hz/s$) & -0.01 & 0.003 & 3\\
rms Amplitude (\%) & 10 & 2 & 20\\
Braking index & 2.5 & 0.5 & 3\\
\hline
\end{tabular}
\caption{Summary of grid values used for signal injection in our simulated time-series. Here $\nu$ and $\dot{\nu}$ are the spin frequency and the spin frequency derivative of the injected signal.}
\label{tab:grid}
\end{table}

Since we are working under the assumption that our deceleration is constant, we also investigated the effect of the braking index n, by setting it to zero while exploring the other grid parameters. The simulations show that the effect of a varying braking index on the detection sensitivity is small. This is indeed expected, since the total length of the observations is short and thus the variation of $\dot{\nu}$ is not dominant. 

\begin{figure}
\centering
\rotatebox{-90}{\includegraphics[width=.33\textwidth]{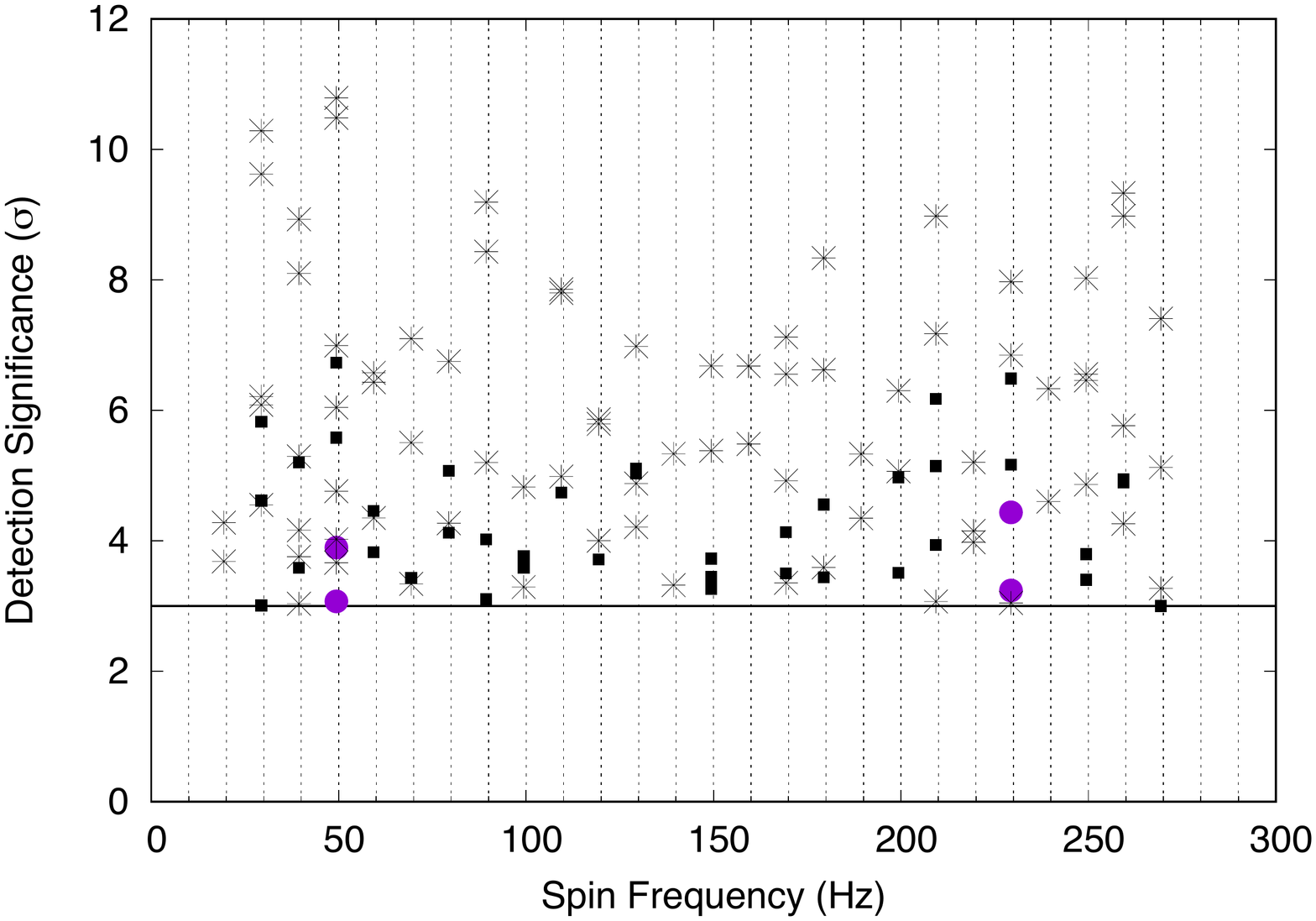}}
\caption{Results of the Monte Carlo simulations on GRB 090515. The vertical grid lines refer to the spin frequency signal injected in the different simulations. The horizontal line is the 3$\sigma$ threshold for a detection. The purple circles are injections with an rms amplitude of 14\%, the black squares 21\% and the gray asterisks 28\%. All injected frequencies (except the 10 Hz one) are recovered with a significance $>3\sigma$ when the rms amplitude is equal to 28\%. Below the 14\% rms amplitude there are no detections above the 3$\sigma$ level. Very similar results are recovered for GRB 090510}
\label{fig:ul}
\end{figure}

For both GRB090515 and GRB090510, the deceleration search shows a robust detection ($>3\sigma$) when the the rms amplitude of the pulsations is larger than about 30\% rms in almost all grid points. The minimum rms amplitude for which we have a detection is about 15\% rms. This means that if a signal of 30\% rms or more had been present in one of the two GRBs analyzed, we would have certainly detected a signal at any of the frequencies accessible. We summarize the results in Figure~\ref{fig:ul}. 

\section{Discussion of possible origins of pulsations}

In the previous sections, we assumed that there would be a periodic signal associated with the spin frequency of a rapidly spinning down magnetar in the lightcurve. We showed that this periodic signal would be detectable in the WT mode observations when using a deceleration search and found no periodic signal in excess of $\approx15$--$30\%$ rms of the total flux. In this section, we discuss the potential sources of periodic emission and the likelihood that they would be detectable.

Our first consideration is the environment of the magnetar. If it is surrounded by an optically thick cloud of material, the fractional amplitude of the pulsed emission drops exponentially with optical depth and, hence, very difficult to detect. During the merger process, a very dense ejecta is expected and this has been modeled in simulations. The ejecta is not isotropic and with a preferred direction along the equatorial plane \citep[e.g.][]{rosswog1999}, leaving the region along the rotation axis reasonably clean. As we are observing emission from the relativistic jet, we know the viewing angle is close to the rotational axis, while the relativistic jet itself is optically thin \citep[e.g.][]{piran2004,metzger2011}. So our viewing angle is most favourable for the periodic emission to escape.

\subsection{Quasi-periodic emission from disk procession}
The accretion disk around the central object (black hole or magnetar) may become warped via differential precession and the amplitude peaks when the spin axis of the central object is highly misaligned relative to the accretion disk \citep{roland1994,blackman1996,portegies1999,reynoso2006, liu2010, stone2013}. The initial quasi-periodic signals have spin periods of the order of 50 ms \citep{stone2013}. However, for SGRBs the accretion disk is expected to be gone within a few seconds \citep{rezzolla2011} and hence this signal would only be expected during the prompt emission and not during our observations so we rule out this mechanism for our analysis. This signal has been sought in periodicity searches of a set of BAT, GBM and BATSE SGRBs by \cite{dichiara2013} but remains undetected to date.

\subsection{Pulsar emission}
The magnetar central engine is a highly magnetised, millisecond pulsar so we might expect it to emit pulses similar to those observed from known pulsars and magnetars, assuming that the observed emission originates directly from the magnetar. We observe periodic emission from pulsars due to a misalignment between the magnetic axis and the rotation axis; a hot spot at the magnetic poles sweeps in and out of view as the neutron star rotates giving a characteristic pulse. The maximal signal occurs when the magnetic axis is orthogonal to the rotation axis and the viewer is also orthogonal to the rotation axis. However, as we have observed a SGRB, we know that the viewing angle is along the initial jet and, hence, close to the rotation axis so very little pulsed emission is expected. There is a chance that the observing axis is off the rotation axis as the jet has a particular opening angle \citep[predicted to be 1--20 degrees;][]{popham1999, ruffert1999, aloy2005, rosswog2005, rezzolla2011}, so there may still be a periodic component to the emission.

However, the magnetic fields and rotation axis are also expected to be highly aligned due to the dynamo mechanism that produces the high magnetic fields \citep{cheng2014, giacomazzo2014}. \cite{cutler2002} show that the rotation axis and dipole field can become orthogonal on a given timescale, the dissipation timescale, if this is less than the electromagnetic spindown timescale (i.e. $<10^{3}T_{em,3}$ s). The dissipation timescale is defined by \cite{cutler2002} as:
\begin{eqnarray}
\frac{1}{\tau_{\rm DIS}} = 3\times10^{-8} {\rm s}^{-1} \left(\frac{10^4}{n}\right) \left(\frac{\nu}{300 ~{\rm Hz}}\right) \left(\frac{\epsilon_{B}}{10^{-7}}\right),   \label{dissipation}
\end{eqnarray}
where n is a factor related to the spin down mechanism and $\epsilon_{B}$ is the quadrupolar distortion of the neutron star due to the magnetic field. This is $\sim 10^7$ s with typical parameters and hence is orders of magnitude longer than the electromagnetic spindown timescales of the magnetars considered in this paper. \cite{dallosso2009} extend this analysis to consider the special case of new born magnetars and show the condition for the two axes to become orthogonal is given by:
\begin{eqnarray}
\frac{E_{B}}{10^{50}~{\rm erg}} \lesssim 2.1 \frac{M_{1.4}}{P_{-3}^2}\left(3+\ln \frac{P_{-3}}{10B_{15}}+\ln \frac{M_{1.4}}{0.48 R_{6}^{4}}\right) \label{dissipation2}
\end{eqnarray}
where $E_{B}$ is the internal magnetic energy. Using typical parameters alongside the predicted magnetic fields and spin periods for the magnetars considered in this analysis, we find the magnetars considered in this paper are typically rotating too slowly for their axes to become completely misaligned on the spindown timescale. Therefore, we expect the magnetic axis and the rotational axis to be close to aligned throughout the spindown timescale \citep[note that even in the case where the rotational axis and magnetic axis are perfectly aligned, the system is still expected to spin down via dipole radiation;][]{goldreich1969}. Observations of known magnetars suggest that the magnetic field and spin axes are typically slightly misaligned \citep{kaspi2017}.

Therefore, very little periodic emission from a pulsar component may be expected due to two reasons:
\begin{enumerate}
    \item The viewing angle is very close to the rotation axis, so only a small proportion of the emission is expected to be pulsed even if the magnetic and rotation axes are completely orthogonal.
    \item The magnetic and rotation axes are likely to be highly aligned at birth and are very unlikely to become orthogonal on the spindown timescales of the magnetars studied in this paper.
\end{enumerate}
In their studies of PSR  J0821-4300, \cite{gotthelf2010} calculated the pulsed fraction of the X-ray emission as a function of the viewing angle and hot-spot angle from the rotation axis. They show that once these angles are greater than  $\simeq$5 degrees, the pulsed X-ray fraction exceeds $\sim$20\% (note there is also an energy band dependence). Therefore, assuming we are directly observing hot-spot emission (similar to that in standard pulsars), our upper limits on the pulsed fraction show that the observing angle and magnetic field axis need to be $\lesssim$5 degrees from the rotation axis.

Considering pulsed emission from magnetars, our limits of 15--30\% are probing many of the typical pulsed fractions observed in known magnetars \citep{israel1999,kargaltsev2012,kaspi2017}. We have used the full energy band of {\it Swift} to obtain sufficient photons, 0.3--10 keV, where the pulsed fraction may be lower \citep{vogel2014}. However, we note that these photons we observe are redshifted and hence we are probing higher energy emission where the pulsed component is expected to be larger.

\subsection{Pulsations from time-dependent scattering in the magnetosphere}
The detection of X-ray pulsations during the radio quiet mode of PSR B0943+10 \citep{hermsen2013} presents an alternative to the standard pulsar model described in Section 4.2. During the radio quiet mode, the X-ray data has a 100\% pulsed thermal component in addition to a non-thermal component, consisting of $\sim 50$\% of the total X-ray emission. PSR B0943+10 has a rotation axis which is thought to be only 9 degrees away from the observer angle and has a nearly aligned magnetic axis, similar to the expected configuration for the magnetar central engine model. \cite{hermsen2013} suggest that the X-ray pulsations originate from a scattered component from within closed magnetic field lines. This model could also be applicable to the magnetars considered in this analysis and, assuming we are able to directly observe the pulsar magnetosphere, may lead to a detectable pulsation signal during the X-ray plateau. 
As we rule out a pulsed fraction of $\sim$15--30\%,  we are not directly observing this emission.

\subsection{Electron acceleration along field lines}
\cite{terada2008} detected pulsations in the hard X-ray component of the emission from the magnetised white dwarf AE Aquarii. The authors propose that the rotating magnetic white dwarf is accelerating electrons along its magnetic field lines, assuming the surrounding medium is a relatively low density plasma \citep{terada2008}. Although the magnetic fields of the magnetars in this paper are orders of magnitude larger than the white dwarf, a similar mechanism could potentially work in the plasma surrounding the newly born magnetar. However, it is not clear if this mechanism would still produce a periodic component if the rotation and magnetic axes are aligned.

\subsection{Reprocessing of emission}

The observed lightcurves are consistent with the energy originating from the spin-down luminosity of a magnetar \citep{zhang2001, metzger2011} however it is not clear where or how the observed X-ray photons are emitted. The mechanisms discussed in Sections 4.2--4.4 assumed we were directly observing the magnetar or its immediate surroundings, however this is unlikely and the emission is most likely to be reprocessed. The magnetar central engine is expected to emit a strong wind that interacts with itself and the local environment \citep[e.g.][]{metzger2011}. This magnetar wind could produce the observed emission via magnetic reconnection or shocks and we consider the likelihood that a periodic signal, from one of the mechanisms outlined earlier, could be retained after reprocessing via these mechanisms:

\begin{enumerate}
    \item {\bf Direct energy injection via forced reconnection:} \\ This theory was originally proposed to explain emission observed in the Crab nebula \citep{lyubarsky2003}. The neutron star emits a magnetised wind which interacts with a surrounding nebula giving a shock at $\sim 10^{17}$ cm. As the neutron star rotates the magnetic field within the wind alternates so, when it reaches the shock front, magnetic reconnection occurs. However, the alternating magnetic fields may not be present due to the alignment of the magnetic field and rotation axis (as discussed in Section 4.2). This model is comparable to models proposed for the prompt emission of GRBs via turbulent magnetic reconnection when there is a collision between two shells with differing magnetic fields \citep[e.g.][]{zhang2011, metzger2011} and would occur at $\sim 10^{15}$--$10^{16}$ cm. This model is consistent with the steep decay phase observed in some lightcurves, as the magnetar wind will stop rapidly when the magnetar collapses to form a black hole. The model proposed by \cite{zhang2011} suggests there will be two variability timescales, one from the central engine and the second from random relativistic turbulence within the emitting regions. In this model there would be some imprint of the millisecond periodic signal from the central engine assuming that the magnetic and rotation axes are misaligned. However, this is likely to be on similar timescales to the relativistic turbulence and hence only constitute a small percentage of the observed signal, making it undetectable in our observations. Even if it is present, a signal of this size would be extremely difficult to detect with current X-ray facilities.
    \item {\bf Direct energy injection via up-scattering of photons in the forward shock:} \\ There is a continued outflow from the central engine, e.g. a magnetar wind, which up-scatters the synchrotron photons left behind the forward shock \citep{panaitescu2008}. If there is a pulsed component in the magnetar wind, due to misaligned magnetic and rotation axes, this could potentially cause a periodic up-scattering of the photons but is likely to be cancelled out due to the variability timescales in the forward shock only being weakly dependent on the input signal timescale \citep{sari1996}. This emission is predicted to occur at $\sim 10^{16}$--$10^{17}$ cm and, if the incoming electrons are hot, this can lead to a scattered signal which is significantly higher luminosity than the standard forward shock emission. The model can explain rapid steep decay phases after the plateau if the scattering outflow suddenly decreases significantly, consistent with the magnetar wind rapidly switching off as the source collapses to a black hole. However, this signal is expected to be brighter for a wind environment, which is not expected for SGRBs. Another disadvantage of this theory for SGRBs is the expectation that SGRBs occur in a very low density environment, hence the forward shock component is expected to be faint - i.e. few photons are available to be up-scattered.
    \item {\bf Indirect energy injection via a refreshed forward shock:} \\ In this scenario, the energy from the magnetar wind is injected directly into the forward shock and hence contributes to the standard forward shock emission \citep[e.g.][]{dallosso2011}. Therefore, as with the up-scattering mechanism, it is unlikely to retain the periodic component \citep{sari1996}. However, this model cannot explain the steep decay phases sometimes observed when the central engine rapidly stops injecting energy into the system. Additionally, this mechanism requires a standard forward shock, which is expected to be weak for SGRBs occurring in low density environments.
    \item {\bf Indirect energy injection via a reverse shock:} \\ Alternatively energy injection, such as from a magnetar wind, is expected to boost the reverse shock \citep{leventis2014,vaneerten2014}. This model is compatible with the low density environments expected with short GRBs and is capable of explaining the steep decay following the plateau phase \citep{vaneerten2014}. This is a very promising mechanism as it is consistent with the observed emission properties and the magnetar central engine model. Unfortunately, as with the forward shock, this mechanism of reprocessing the emission would most likely obliterate any periodic component in the energy injection.
\end{enumerate}

All these are viable emission mechanisms within the magnetar central engine model, but only the forced reconnection model holds the potential of retaining some of the underlying temporal structure from the central engine, however the periodicities we are searching for are comparable to the random reconnection timescales. Therefore, with the sensitivity of current X-ray facilities, if the magnetar emission has been reprocessed we are very unlikely to be able to extract a periodic signal from the random noise component. 

\section{Conclusions}

Plateaus in the X-ray lightcurves of short GRBs are signatures of energy injection that are thought to originate from a newly formed magnetars rapidly spinning down due to the emission of dipole radiation. Using the magnetar central engine model, we are able to predict the spin-down frequency and rate that may result in an evolving periodic component in the observed X-ray emission (similar to that observed in pulsars and Galactic magnetars). In this paper, we show that the frequency of the periodic component is detectable within the capabilities of the WT mode of the XRT onboard the {\it Swift} Satellite and calculate the optimal parameter space to search for 2 SGRBs.

We have conducted a deceleration search for a periodic signal during X-ray plateaus following these SGRBs and, taking into account rapid spin-down via dipole radiation, do not detect any periodic component to a limit of $\approx$15--30\% rms. The rotation and magnetic axes of the magnetar are likely to be close to alignment, unfavourable for the production of a significant periodic component. We show that this signal is still potentially attainable if we are directly observing emission from the magnetar central engine. However, the emission is likely to be reprocessed by the magnetar wind interacting with the forward or reverse shocks and the reprocessing mechanisms are likely to reduce any periodic component to a few percent of the total emission. With future, more sensitive instrumentation \citep[e.g. enabling us to check for energy dependencies in periodic emission;][]{vogel2014} and more complex search models it will be possible to place much more stringent limits on the presence of a periodic component or lead to a detection that would confirm the magnetar central engine model.

\section{Acknowledgements}
We are grateful to Phil Uttley for his suggestion to look into this and to Yuri Cavecchi, Daniela Huppenkothen, Anna Watts and Ralph Wijers for very useful discussions. AP acknowledges support from a Netherlands Organization for Scientific Research (NWO) Vidi Fellowship. PTO would like to acknowledge funding from STFC. This work makes use of data supplied by the UK {\it Swift} Science Data Centre at the University of Leicester and the {\it Swift} satellite. {\it Swift}, launched in November 2004, is a NASA mission in partnership with the Italian Space Agency and the UK Space Agency. Swift is managed by NASA Goddard. Penn State University controls science and flight operations from the Mission Operations Center in University Park, Pennsylvania. Los Alamos National Laboratory provides gamma-ray imaging analysis.

\end{document}